\documentstyle[12pt,aaspp4]{article}
\input epsf
\begin{document}
\title{Cluster Abundance Constraints on Quintessence Models}
\author{Limin Wang and Paul J. Steinhardt}
\affil{Department of Physics and Astronomy,
University of Pennsylvania,
Philadelphia, Pennsylvania 19104 USA}

\begin{abstract}
The abundance of rich clusters is a strong constraint on the mass
power spectrum. The current constraint can be expressed in the form
$\sigma_8 \Omega_m^{\gamma} = 0.5 \pm 0.1$ where $\sigma_8$
is the $rms$
mass fluctuation on 8~$h^{-1}$~Mpc scales, $\Omega_m$ is the
ratio of matter density to the critical density, and $\gamma$ is 
model-dependent. In this paper, we determine a general expression
for $\gamma$ that applies to any models with a mixture of cold 
dark matter plus cosmological constant or quintessence (a 
time-evolving, spatially-inhomogeneous component with negative 
pressure) including dependence on the spectral index $n$,
the Hubble constant $h$, and the equation-of-state of the quintessence
component $w$. The cluster constraint  is combined 
with  COBE  measurements to identify a spectrum of best-fitting models. 
The constraint from the evolution of rich clusters  is also discussed.

\end{abstract}
{\it Subject headings:}
  cosmology: theory - dark matter - large-scale structure of universe
 -  galaxies: clusters: general - X-rays: galaxies

\newpage

\section{Introduction:}

Rich clusters can be used to constrain cosmological models of 
large-scale structure formation.  Rich clusters are the largest
virialized objects in the universe and, hence, their abundance and
evolution can be simply related to the linear mass power spectrum,
$P(k)$.  Their x-ray temperature can be used to infer the cluster mass.
Then, Press-Schechter (1974) theory can be used to  relate the 
observed cluster abundance as a function of mass 
to the $rms$ fluctuation on 
8~$h^{-1}$~Mpc scales, $\sigma_8$.  The result is a constraint on
a combination of parameters,
 $\sigma_8 \Omega_m^{\gamma}$, where $\gamma$
is a function of model parameters.

In this paper, we obtain a general expression for $\gamma$ 
 which applies to a wide range of
models including the standard cold dark matter (sCDM) model, 
models with a mixture of cosmological constant ($\Lambda$) and 
cold dark matter ($\Lambda$CDM), and QCDM models with a mixture of 
cold dark matter and quintessence, a dynamical, time-evolving,
spatially inhomogeneous component with negative pressure.
An example of  quintessence would be a scalar field ($Q$)
 rolling down a potential $V(Q)$.  
It has been shown that a spectrum
of $\Lambda$CDM and QCDM models satisfy all current observational
constraints (Wang et al. 1998). 
Our expression for $\gamma$ includes dependence on the spectral index $n$,
the Hubble constant $h$ (in units of 100 km sec$^{-1}$ Mpc$^{-1}$), and the 
equation-of-state of the quintessence component $w$.

In Section 2, we discuss our derivation of the cluster abundance 
constraint, which follows earlier derivations to some degree but 
includes some new features when applied to quintessence.
Some may wish to proceed directly
to the resulting constraint, given in Section 3.
Our result appears to differ slightly  from some earlier works, but we 
explain in this section the reasons for  those differences.
Our purpose in determining $\gamma$ is to transform observations 
into a powerful constraint on models.  
In Section 4, we discuss the further constraint derived from studying 
the evolution of cluster abundance from redshift $z=0$ to $z \sim 1$.
In Section 5, we conclude by  applying 
the cluster abundance constraint on $\sigma_8$ in combination 
with the constraint on  COBE normalization of 
$P(k)$ to pick out a spectrum of best-fitting $\Lambda$CDM and 
QCDM models.

\section{Derivation of  the Cluster Abundance Constraint for
Quintessence Models}

In this section, we present the derivation of the cluster abundance 
constraint on $\sigma_8$ for QCDM models. We take a pedagogical
approach in which we first discuss each step for a $\Lambda$CDM 
model and then point out the differences that arise in QCDM models.
The final result can be found in Section 3.

\subsection{Mass-temperature relation} 
Press-Schechter theory relates the number density of 
clusters to their mass.  Observations
determine directly 
 the temperature instead of the mass of clusters.
Therefore,  the mass-temperature relation
derived from 
the virial theorem (Lahav et al. 1991; Lilje 1992)  is used
to apply the Press-Schechter relation.

First, let's consider a universe
with vacuum energy density $\rho_{\Lambda}$. For  a
spherical overdensity, we have
\begin{equation} \label{virial}
T_{vir} = -\frac 1 2 U_G + U_{\Lambda}
\end{equation}
where $T_{vir}$ is the kinetic energy, $U_G$ is the potential energy 
associated with the spherical mass overdensity, and $U_{\Lambda}$ is
the  potential energy associated with $\Lambda$.
The kinetic energy is $T_{vir}=\frac 1 2 M \bar{V_{vir}^2}$ where
$M$ is the mass of the cluster and $\bar{V_{vir}^2}$ is the mean square 
velocity of particles in the cluster. 
The gravitational
potential is $U_G =  - \frac 3 5 \frac{G M^2}{R}$ where $G$ is the
gravitational constant and $R$ is the radius of the cluster.   The 
potential due to vacuum energy is 
$U_{\Lambda} = - \frac 4 5 \pi G \rho_{\Lambda}  M R^2$.  
Re-expressing the background matter energy density at redshift $z$ as
$\rho_b=\frac{3H_0^2}{8\pi G}\Omega_0 (1+z)^3$ where
$H_0$ the is the  Hubble parameter, $\Omega_0$ is the ratio 
of the matter density to the critical density, 
 and $z$ is the redshift,  
the virial relation (\ref{virial}) becomes:
\begin{equation} \label{v_vir}
\bar{V_{vir}^2}=\frac 3 5 \left(G M H_0 \sqrt{\Omega_0 \Delta_c/2}
\right)^{2/3}
(1+z)\left(1-\frac 2 {\Delta_c}\frac{\Omega_{\Lambda}(z)}
{\Omega_{m}(z)}\right)
\end{equation}
where $\Delta_c \equiv \rho_{cluster} / \rho_{b}$ is the ratio of 
the cluster to background density, and
 $\Omega_{\Lambda}(z)$ and 
$\Omega_{m}(z)$ are the density parameters for vacuum energy and
matter at redshift $z$, respectively.
The observed temperature of the gas	 is
\begin{equation} \label{T_vir}
k_B T=\frac{\mu m_p}{\beta}\sigma_{los}^2=\frac{\mu m_p}{\beta}
\frac{\bar{V_{vir}^2}}{3}
\end{equation}
where $\mu m_p$ is the mean mass of particles;
$\sigma_{los}^2$ is the line-of-sight velocity dispersion;
$\beta$ is the ratio of the kinetic energy to the temperature
and $k_B$ is the Boltzmann constant.
The mass of the cluster is then:
\begin{equation} \label{M_vir}
M=\left(\frac{k_B T}{0.944 f_{\beta} \, {\rm keV}}\right)^{3/2}
\left[(1+z)^3\Omega_0 \Delta_c\right]^{-1/2}
\left[1-\frac{2}{\Delta_c}
\frac{\Omega_{\Lambda}(z)}{\Omega_{m}(z)}\right]^{-3/2} 
h^{-1}10^{15} M_{\odot}
\end{equation}
where $f_{\beta} \equiv f_u \mu /\beta$ 
where $f_u$ is a fudge factor of order unity that allows deviation from the
simplistic spherical model.
Using this relation,
the mass-temperature relation of a virialized
cluster can be computed at any $z$.  However,
if we evaluate at redshift $z=z_c$, twice the
turn-around time  (that is,
 $t(z_c)=2t(z_{ta})$ where $z_{ta}$
is the redshift at which the cluster turns around,
 $\dot{R}=0$),
then $\Delta_c(z_c,\Omega_m)=\Delta_c(\Omega_m)$ becomes a
function of $\Omega_m$ only (for $\Lambda$ and open universes).
Note that
$z_c$ is  the redshift at which
the cluster formally collapses to $R=0$ according to
an unperturbed spherical solution.

In quintessence models, the principal difference is that 
the energy density in $Q$ decreases with time, whereas vacuum 
energy remains constant.
The  $Q$-component 
 does not cluster on scales less than 100~Mpc 
(Caldwell, Dave, \& Steinhardt 1998).
Consequently, the only  effect of $Q$ on the abundance of rich clusters
with size less  than 100 Mpc 
 is through its modification of  the background
evolution.  We will restrict ourselves to cases where
the equation-of-state $w$ is constant or slowly varying.  In this case,
the ratio  
$\Omega_{\Lambda}(z)/\Omega_m(z)$ above can be replaced by
$(1+z)^{3 w}\Omega_Q(z=0)/\Omega_0$.

For quintessence models, the ratio of cluster to background density, 
$\Delta_c|_{z_c}=\Delta_c(\Omega_m,w)$, is a  function of two variables:
\begin{equation} \label{Delta_c}
\Delta_c(z=z_c) \equiv \frac{\rho_{cluster}(z_c)}{\rho_{b}(z_c)} =
\zeta\left(\frac{R_{ta}}{R_{vir}}\right)^3
\left(\frac{1+z_{ta}}{1+z_c}\right)^3.
\end{equation}
where $\zeta \equiv \rho_{cluster}(z_{ta})/\rho_b(z_{ta})$; 
$R_{ta}$ and
$R_{vir}$ are the radius of the cluster at $z=z_{ta}$ and 
at  virialization, respectively. (The second equality utilizes
the standard assumption
 that  the cluster has  virialized at $z=z_c$.)
The factor $\zeta$ has been computed by solving for the evolution
of a spherical overdensity in a cosmological model with constant $w$
(see Appendix A).  
We find that $ \zeta$ is weakly model-dependent:
 $\zeta= (3\pi /4)^2 \Omega_m ^{-0.79+0.26\Omega_m-0.06w}|_{z_{ta}}$ 
for $-1 \le w \le 0$.  
The fact that this expression is weakly model-dependent means that
we can also apply it to models with time-varying $w$.

By the virial theorem and energy conservation, we have:
\begin{equation}  \label{U}
\frac 1 2 U_{G}(z_c) + 2U_{Q}(z_c)=U_{G}(z_{ta})
+U_{Q}(z_{ta})
\end{equation}
This leads to an approximate solution
\begin{equation}  \label{r_vOr_t}
\frac{R_{vir}}{R_{ta}}=\frac{1-\eta_v/2}{2+\eta_t-3\eta_v/2}
\end{equation}
where $\eta_t= 2\zeta^{-1}\Omega_{Q}(z_{ta})/\Omega_m(z_{ta})$
and $\eta_v=2\zeta^{-1}((1+z_c)/(1+z_{ta}))^3
 \Omega_{Q}(z_c)/\Omega_{m}(z_c)$.
The fact that $\eta_v$ is time-dependent if $w\neq -1$ is  
because the energy within the cluster is not exactly conserved and
the temperature is shifting after virialization due to the
change in the $Q$-energy, $\rho_Q$, within the cluster.  However, since the
matter density $\rho_m$ is much larger than $\rho_Q$ in 
a collapsed cluster, this correction is negligible.

\subsection{Density Dispersion} 
The dispersion of the density field on a given comoving scale $R$ is
\begin{equation} \label{sigmaR}
\sigma^2(R,z)=\int^{\infty}_0{W^2(kR)\Delta^2(k,z)\frac{d\, k}{k}}
\end{equation}
where
\begin{equation} \label{window}
W(kR)=3\left(\frac{sin(kR)}{(kR)^3}-\frac{cos(kR)}{(kR)^2}\right)r,.
\end{equation}
and
\begin{equation} \label{Delta}
\Delta^2(k,z)=4\pi k^3 P_{\delta}(k,z).
\end{equation}
$P_{\delta}(k)\equiv |\delta_k|^2$ is the power spectrum and 
$\delta_k$ is the Fourier transform of the fractional density
perturbation $(\delta\rho/\rho)^2\equiv\langle \delta^2(\vec{x})\rangle$
\begin{equation} \label{delta_k}
\delta_k\equiv \delta(|\vec{k}|) = \frac{1}{(2\pi)^{3/2}}
\int{d^3x \delta(\vec{x})e^{i\vec{k}\cdot
\vec{x}}}.
\end{equation}
For constant or slowly varying $w < -1/3$, the BBKS approximation 
to the power spectrum (Bardeen et al. 1986) is reliable. However, 
if $w > -1/3$ or if $w$ is  rapidly varying, we find no general
fitting formula; instead,  
$P(k)$ must be obtained   numerically.

The $rms$ mass fluctuation 
$\sigma_8(z) \equiv \sigma(R=8h^{-1}Mpc,z)$ can be expressed as
$\sigma_8(z)=g(z) \sigma_8(z=0)$, where $g(z)$ is the
growth factor.  The growth factor is
  proportional to the linear density perturbation
$\delta_{\rho}\equiv \delta \rho/\rho$ and normalized to $g(z=0)=1$. 
We find that a   
good approximation to the the growth index is given by (see Appendix B)
\begin{equation} \label{f}
f_g\equiv \frac{d\, \ln \delta_{\rho}}{d\, \ln a} = \Omega_m(z)^{\alpha}
\end{equation}
where $a=1/(1+z)$ is the scale factor and
\begin{equation} \label{alpha}
\alpha=\frac{3}{5-\frac{w}{1-w}}
+\frac{3}{125}\frac{(1-w)(1-3w/2)}{(1-6w/5)^3}(1-\Omega_m) + {\cal O}((1-\Omega_m)^2)
\end{equation}
The growth factor $g(z)$ is obtained from the integral expression
\begin{equation} \label{g}
g(z) \approx a\,\,{\rm exp} \, \left[ \int_{a}^1 \frac{da}{a} (1-\Omega_m^{\alpha} ) \right]
\end{equation}
We tested
the expression for $g$ obtained from Eq.~(\ref{alpha}) 
against the value obtained
by numerically integrating the density perturbation equations; for
$1-\Omega_m$ between zero and 0.8,
the accuracy is better than 1\%.

\subsection{Press-Schechter Theory} 
According to Press-Shechter theory,
 the comoving number density
of virialized objects with mass $M=4\pi R^3\rho_b/3$ 
 is:
\begin{equation} \label{PS}
d\, n(M,z)=-\sqrt{\frac 2 {\pi}}\frac{\rho_b}{M^2}
\frac{\delta_c R}{3\sigma^2(R,z)}\frac{d\, \sigma(R,z)}{d\, R}
\exp\left[-\frac{\delta_c^2}{2\sigma^2(R,z)}\right]d\, M
\end{equation}
where  $\delta_c=\rho_{linear}/\rho_b$
is the perturbaton  that would be predicted for a spherical 
overdensity of radius $R$ and mass $M$ according to linear 
 theory.
Given the observed number density $\Delta n$ 
within a certain temperature
range $\Delta T$, 
Eqs.~(\ref{M_vir}) and (\ref{PS}) 
can be used to  determine the normalization of
the mass power spectrum $\sigma_8$. 

The
major uncertainties in this method are the observational
error in the number density $\Delta n$ and the systematic
error in determining the model parameters $\delta_c$
in Eq.~(\ref{PS}) and $f_{\beta}$ 
in Eq.~(\ref{M_vir}).
Specifically
\begin{equation} \label{error}
\frac{\delta \sigma_8}{\sigma_8}=
u_1\frac{\delta \Delta n}{\Delta n} 
+\frac{\delta (\delta_c)}{\delta_c}
-\left[\left(1+\frac{u_2+u_3}2\right)u_1 - \frac{u_2}2  \right]
\frac{\delta f_{\beta}}{f_{\beta}}
\end{equation}
where $u_1=[\delta_c^2/\sigma(R)^2 -1]^{-1}$,
$u_2=-d\ln{\sigma(R)}/d\ln{R}$ and 
$u_3=-d\ln{\sigma(R)'}/d\ln{R}$
are positive
and of order unity
in  the range of interest.
By studying spherical models, we find that $\delta_c$ varies
slowly as a function of $\Omega_m$, 
$1.6 < \delta_c < 1.686$. We also find that
$f_{\beta}$ does not depend on the cosmological 
model and can be determined by numerical simulation. 

According to Eq.~(\ref{M_vir}), the virial
temperature corresponding to a given virial mass depends on 
the redshift at which a cluster is virialized.  Therefore, to get the number
density of clusters of a given temperature range today,
we need to find out the virialization rate and integrate from
$z=0$ to $z=\infty$.  Assuming that the merger of
clusters is negligible, the Press-Schechter relation, Eq.~(\ref{PS})
can be re-expressed as:\\
\begin{equation} \label{dPS}
\frac{d\, n(T,z)}{dz}=-\sqrt{\frac 1 {2\pi}}\frac{\rho_b}{M(T,z) T}
\frac{d\, \ln\sigma(R,z)}{d\, \ln R} 
\frac{d\, \ln \sigma_8(z)}{dz}
x(x^2-1)e^{-\frac{x^2}{2}}d\, T
\end{equation}
where $x=\delta_c/\sigma(R,z)$.

Lacey \& Cole (1993,1994) and Sasaki (1994)
have estimated the corrections
due to  cluster merging 
(see also, Viana \& Liddle 1996).
The  corrections are small.
In our results, we  average the two  estimates of the merging
correction to obtain our final result.

\section{Generalized Cluster Abundance Constraint}
The cluster abundance constraint on $\sigma_8$ 
is obtained by comparing
the theoretical prediction discussed in the previous section
to observations.
The observed X-ray cluster abundance as a function 
of temperature was presented
by Edge et al (1990) and Henry \& Arnaud (1991, hereafter HA).  
After a recent correction (Henry 1997) to HA, the two results agree.
  We have 
fit the theoretically predicted number density {\it vs.} temperature
curve (the temperature function) 
to the HA data.  

\subsection{Principal Results}
 Our results can be
fit by
\begin{equation} \label{sigma8}
\sigma_8=(0.50-0.1\Theta)\Omega^{-\gamma (\Omega,\Theta)}
\end{equation}
where $\Theta\equiv \Theta_n + \Theta_h 
=(n_s-1)+(h-0.65)$ where $n_s$ is the spectral index of
primordial energy  density perturbations and $h$ is  the 
present Hubble constant in units of $100 {\rm km \,  s^{-1} \, Mpc^{-1}}$. 
For QCDM models with equation-of-state $w$ (including
$\Lambda$CDM with  $w=-1$), our fit to $\gamma$ is
\begin{equation} \label{QCDMf}
\gamma (\Omega,\Theta)=0.21-0.22w+0.33\Omega+0.25\Theta.
\end{equation}
For open models
\begin{equation} \label{openf}
\gamma (\Omega,\Theta)=0.33+0.35\Omega+0.20\Theta.
\end{equation}

\subsection{Comparison to Previous Computations}
Many groups have presented   similar constraints on $\sigma_8$  for $\Lambda$CDM
and OCDM models. In general, all of them are in reasonable agreement
with one another and with our result. 
We identify below the sources of 
the  discrepancies, some real and some only apparent, when compared to 
White, Efstatiou, \& Frenk (1993, hereafter WEF);
Eke, Cole, \& Frenk (1996, hereafter ECF); 
Viana and Liddle (1996, hereafter VL);
Pen (1997, hereafter Pen); 
Kitayama and Suto (1997, hereafter KS).

\noindent (1) As we argued, an integration (with merger correction)
of Eq.~(\ref{dPS}) is necessary since the mass-temperature 
relation is redshift dependent.  
Most groups only applied
Eqs.~(\ref{M_vir}) and (\ref{PS})
at redshift $z=0$ to obtain their main
results, which  leads to an 
overestimate of $\sigma_8$  by as much as $10\%$. However, some groups
(ECF and Pen) fit the number density and temperature relations
to  numerical simulations to normalize their
coefficient $f_{\beta}$ in Eq.~(\ref{M_vir}).  
In so doing, they effectively incorporated the integration 
correction into the coefficient $f_{\beta}$ for the cases that
were numerically tested. 
Consequently, their fitted value of  $f_{\beta}$
does not represent precisely $f_u \mu/\beta$ 
as defined in Eq.~(\ref{M_vir}).
 We shall call their corresponding
coefficient $\tilde{f}_{\beta}$ to emphasize that the physical meaning
of this coefficient has been modified to include intregration over
redshift.
Since the contribution of the integration is not proportional
to  $f_{\beta}$ 
in all cosmological models,
it is more precise and physically meaningful to do separately
the redshift integration.

\noindent (2) The shape of the theoretical
cluster temperature function,
Eq.~(\ref{dPS}),
does not agree equally well with  observations 
for all parameters. The fit is particularly poor for
  models with large 
$\Gamma \equiv \Omega_m h$ and positive tilt ($n_s>1$).
To handle this problem,
some groups (WEF; Pen; VL)
only fit  the observed number density at one particular 
 temperature;  this 
introduces  some arbitrariness and leads to much larger 
uncertainties depending on which 
temperature is chosen.  In our analysis,
we fit the theoretical temperature 
function to all the data points provided by HA.  

\noindent (3)  The 
recent correction to the HA data results in a correction to the
VL results of 20\%.

\noindent (4) Most groups (WEF; ECF; VL; Pen) assume a fixed 
``shape parameter'' $\Gamma = \Omega_m h$. We found that  no
single 
$\Gamma$ is valid for all QCDM models.
Instead of expressing our results in terms of fixed $\Gamma$,
we fix 
 $h=0.65$.    We  include the dependence on $h$
($\Theta_h$) explicitly.

\noindent (5) Most recent analyses (ECF; VL; Pen; KS)
adopted similar modeling
of the mass-temperature relation.  
However, there is still
about 10\% to 20\% disagreement on the value of
$f_{\beta}$ in Eq.~(\ref{M_vir}) due to the uncertainties
of the numerical simulations.  We found that the most extensive simulation
results, those  of ECF and Pen, agree very well with each other.
By normalizing our theoretical calculations to their 
simulations, we found $f_{\beta}=1.1$. Notice that their
reported  values for the coefficient are $\tilde{f}_{\beta}=1.4$  
and $\tilde{f}_{\beta}=1.3$, respectively. This discrepancy
is due to rolling into $\tilde{f}_{\beta}$ the integration effect
described under (1).  Recent simulations by Bryan and 
Norman (1997) also indicate a similar
result, once one corrects for their slightly higher value of 
 $\delta_c$.

\noindent (5) Some groups (VL) used 
the differential temperature function 
(the cluster abundance within unit temperature interval
around a center value $T_{cen}$) 
while others (WEF; ECF; Penn; KS) used cumulative temperature function
(the cluster abundance with temperature above a critical
temperature $T_{cri}$).
These two approaches give similar results because
the cluster abundance drops exponentially with temperature
and the cumulative cluster abundance is well approximated
by counting the cluster abundance around $T_{cri}$.
We compared the results obtained by using 
the differential temperature function given by HA and
that obtained by using the cumulative temperature function
provided by ECF and found
them to be in  good agreement.  However, the
error bar of the latter is much smaller: most models were
excluded by 95\% confidence level by the temperature function
fitting.  To be conservative, we used the former to get our
results and errors.

\subsection{Error Estimates}
 Eq.~(\ref{error})
 can be used to estimate the total error for 
$\sigma_8$. 
 From the present scatter of numerical 
simulation results, $f_{\beta}$ has about 20\%
uncertainty, $\delta_c$ has about 10\% uncertainty and
another 15\% uncertainty comes from the observation.
Therefore, the net uncertainty 
quoted  in Eq.~(\ref{sigma8})
is  about 20\% 
corresponding to 95\% confidence level.

\section{Evolution of Abundance}
By applying the same theoretical tools, we can also study the evolution
of the cluster abundance
to obtain further constraints on $\sigma_8$ and $\Omega_m$.  The current
redshift survey results were converted to number densities of clusters
with their comoving-1.5 mass (the mass within comoving radius 
$R_{com}=1.5 h^{-1}Mpc$) greater than a given  mass 
threshold $M_{1.5}$ (Carlberg et al. 1997; Bahcall, Fan, \& Cen 1997).  
If the mass profile
for the cluster obeys $M(R) \propto R^p$ near $R=1.5 h^{-1}Mpc$, 
and the average virial overdensity
is equal to $\Delta_c$ as calculated in Eq.~(\ref{Delta_c}), 
then the virial mass $M$ is related to $M_{1.5}$ by 
\begin{equation}  \label{M1.5}
M=\left(\frac{178}{\Delta_c}
\frac{M_{1.5}}{6.97\times 10^{14}\Omega_0 h^{-1} M_{\odot}}
\right)^{\frac{p}{3-p}}M_{1.5}
\end{equation}
 Eq.~(\ref{PS}) can be used to estimate the number density
of observed objects at a given redshift.  We adopted $p=0.64$
as suggested by Carlberg, Yee, \& Ellingson (1997).

The log-abundance as a function of $z$, $\log_{10}{\left(n(M_{1.5},z)\right)}$,
is roughly linear as a function of $z$ for $0<z<1$ for the models of 
interest.
A useful parameter to characterize
 the evolution of cluster abundance at redshift 
$0 \le z \le 1$ is $A(M_{1.5})$, defined by 
\begin{equation} \label{n-z}
\log_{10}{\left(n(M_{1.5},z)\right)} \approx A(M_{1.5}) z +B(M_{1.5})
\end{equation}
where $n(M_{1.5},z)$ is the number density of clusters with
comoving-1.5 mass greater than $M_{1.5}$ observed at redshift $z$.
The smaller $A(M_{1.5})$ is, the stronger the evolution is.
By applying this analysis to models, we found \\
(1) 
Cluster abundance
evolution strongly depends on $\sigma_8$: low $\sigma_8$ 
leads to strong evolution. This agrees
with what Bahcall, Fan, \& Cen (1997) have found.
This is a general feature of
Gaussian-distributed random density peaks. \\
(2) Cluster abundance is also sensitive to the equation-of-state
of quintessence $w$: low $w$ leads  to strong evolution.
With the same $\Omega_0$, the growth of density perturbations gets 
suppressed earlier in high $w$ models, therefore, they have a weaker
evolution in recent epochs ($0 \le z \le 1$).  \\

In Figure~1, we
show $A(M_{1.5}=8\times 10^{14}h^{-1}Mpc)$ as a function
of $\sigma_8$ for some sample models which have been chosen because
they all fit current observations well (see discussion in following section).
We allow $\sigma_8$ to vary
from 0.5 to 1.0 with the COBE normalized $\sigma_8$ shown 
as opaque circles.  
The current 
redshift survey data (Bahcall \& Fan 1998) indicate
$A(M_{1.5}=8\times 10^{14}h^{-1}Mpc) \gtrsim -5$ at the 3$\sigma$ level
(with mean equal to -1.7), 
 which is consistent with 
all six COBE normalized models. 

\section{Conclusions}

The cluster abundance and evolution constraints, when combined
with future measurements of the cosmic microwave background, 
may be an  effective means of discriminating quintessence
and $\Lambda$ models.

The cosmic microwave background (CMB) anisotropy provides a constraint
on the mass power spectrum on the horizon scale. For a given model,
this constraint from large-scale anisotropy as measured by the COBE-DMR
satellite (Smoot et al. 1992; Bennett et al. 1996) 
can be extrapolated to obtain a limit on 
$\sigma_8$ that is completely independent of the cluster abundance
constraint.  In Figure~2, we plot the dependence of $\sigma_8$
on $\Omega_Q$.  For each $w$, a different curve is shown.  Along
each curve is highlighted the range of $\sigma_8-\Omega_Q$
consistent with  the cluster abundance constraint derived  in this paper.
Hence, the best-fit models are those  near the middle of the 
highlighted regions.  These are the same models used as 
examples  in Figure~1.

Near future satellite experiments, such as the NASA 
Microwave Anisotropy Probe (MAP) and the 
ESA Planck mission, will greatly improve upon COBE by
determining the temperature anisotropy
power spectrum to extremely high precision from  large to small
angular scales.  Even a full-sky, cosmic variance limited measurement
of the CMB anisotropy, though, may  not be sufficient to
 discriminate $\Lambda$CDM
from QCDM models.  There is a degeneracy in parameter space such
that, for any given $\Lambda$CDM models, there is a continuous family 
of QCDM models which predicts the 
same CMB power spectrum (Huey et al. 1998; White 1998)
It is possible that the data points to a QCDM model, say, which lies
outside this degenerate set of models, {\it e.g.}, a model with rapidly
varying $w$. However, if the data  points to the degenerate set of models,
then it is critically important to find a method of discriminating models
further.  Not only does degeneracy mean that $\Lambda$ cannot be 
distinguished from quintessence, but 
also that large uncertainty in $\Omega_m$ and $h$.
Here we wish to illustrate how  cluster abundance may play an important role.

An example of a  ``degeneracy  curve" is shown
in Figure~3.
Given a value of $h$ for any one point along the curve, values of $h$
can be chosen for other points along the curve such that the models
are all indistinguishable from CMB measurements.
The near-future satellites are capable of limiting parameter space
to a single degeneracy curve. The CMB anisotropy also narrowly 
constrains $n_s$, $\Omega_b h^2$, and $\Omega_m h^2$.
However, even when combined with the cluster abundance constraint
and other cosmological constraints from the age, Hubble constant,
baryon fraction, Lyman-$\alpha$ opacity, deceleration parameter and
mass power spectrum, a substantial degeneracy can remain. 
 Figure~3 
includes a shaded region which exemplifies the range which these 
models might allow, 
based on current measurements (Wang et al. 1998; Huey et al. 1998).
Because of the uncertainty in cluster abundance at $z=0$ and
other cosmic 
parameters, the overlap between the degeneracy curve and the shaded region 
allows a wide range of $\Omega_m$, $h$ and $w$. 

Cluster evolution offers a promising approach for breaking the degeneracy.
Figure~4 illustrates the variation of $A(M_{1.5})$ as a function of 
$w$ for models along the degeneracy curve and inside the shaded region
of Figure~3. The variation in $A(M_{1.5})$ is nearly 2, corresponding
to nearly two orders of magnitude variation in abundance at redshift 
$z=0.5$. The range of $A(M_{1.5})$  is between -3.5 and -5.5 in this
case, but this could be shifted upward or downward by adjusting 
cosmic parameters. The point is that models which are degenerate in
terms of CMB anisotropy are spread out in $A(M_{1.5})$.
If the measurements can be refined so that $A(M_{1.5})$
is know to better than $\pm 0.5$, 
then cluster evolution may play an important role in discriminating
between quintessence and  vacuum density and, thereby, determining
$\Omega_m$ and $h$.

\acknowledgements
We have benefitted  greatly from many  discussions with
  N. Bahcall and R. Caldwell.  We also thank
P. Bode, G. Bryan, A. Liddle, Y. Suto and P. Viana for useful suggestions.
This
research was supported by the Department of Energy at Penn, DE-FG02-95ER40893.

\newpage

\appendix
\section{Spherical Model}
\label{AppA}
We study a spherical overdensity with uniform matter density 
$\rho_{cluster}$ and radius $R$ in a
background that satisfies the Friedmann equation:
\begin{equation}\label{background}
\left(\frac{\dot{a}}{a}\right)^2=\frac{8\pi G}{3}(\rho_b + \rho_Q)
\end{equation}
where $a$ is the scale factor, $\rho_b$ is the background matter
energy density and $\rho_Q$ is the energy density in $Q$.
Quintessence doesn't cluster at the interesting scales; 
the energy density in
$Q$ remains the same both inside and outside the overdensity patch.
Because the curvature is not a constant inside the overdensity patch. 
we  use the time-time component of the Einstein equations 
(which does not involve the curvature term)
to solve for the growth of the overdensity patch
\begin{eqnarray}  \label{overdensity}
\frac{\ddot{R}}{R} & = &
-4\pi G\left(p_Q+\frac{\rho_Q+\rho_{cluster}}{3}\right)  \nonumber  \\
&=&-4\pi G\left[\left(w+\frac13\right)\rho_Q + \frac13\rho_{cluster}\right]
\end{eqnarray}
We have used $p_Q=w\rho_Q$ to obtain the second equality.  Now, 
we define 
\begin{eqnarray}
x & \equiv & \frac{a}{a_{ta}} \\
y & \equiv & \frac{R}{R_{ta}}
\end{eqnarray}
where $a_{ta}$ and $R_{ta}$ are the scale factor and the radius
at turn-around time, then
\begin{eqnarray}
\rho_b &=& \frac{3H_{ta}^2}{8\pi G}
	\frac{\Omega_m(x=1)}{x^3} \\
\rho_Q &=&  \frac{1-\Omega_m(x)}{\Omega_m(x)} \rho_b \\
\rho_{cluster}& =& \frac{3H_{ta}^2}{8\pi G}
	\frac{\Omega_m(x=1)}{y^3}\zeta   
\end{eqnarray}
where $\Omega_m(x)$ is the matter energy density parameter at $x$,
$H_{ta}$ is the Hubble constant at turn-around time and 
$\zeta \equiv (\rho_{cluster}/\rho_b)|_{x=1}$.  Eqs.~(\ref{background})
and (\ref{overdensity}) can be then written as
\begin{eqnarray}  \label{x}
\frac{d\, x}{d\, \tau}& = &\left(a \Omega_m(x)\right)^{-1/2} \\
\label{y}
\frac{d^2y}{d\tau^2}& =& -\frac12\left[\frac{\zeta}{y^2}
	+(1+3w)\frac{1-\Omega_m(x)}{\Omega_m(x)}\frac{y}{x^3}\right]
\end{eqnarray}
where $d\tau = H_{ta} \sqrt{\Omega_m(x=1)} dt$. With the boundary 
condition $dy/d\tau|_{x=1} = 0$ and $y|_{x=0}=0$, 
$\zeta$ is uniquely determined 
by Eqs.~(\ref{x}) and (\ref{y}), given the function form of $\Omega_m(x)$.
For constant $w$, we have
\begin{equation} \label{Omega_m}
\Omega_m(x)=\left(1+\frac{1-\Omega_m(x=1)}{\Omega_m(x=1)}x^{-3w}\right)^{-1}
\end{equation}
and $\zeta$ obtained from Eqs.~(\ref{x}) and (\ref{y}) can be well
fitted by 
\begin{equation}  \label{zeta}
\zeta= \left(\frac{3\pi}{4}\right)^2 
\Omega_m ^{-0.79+0.26\Omega_m-0.06w} |_{x=1}
\end{equation}
The linear overdensity $\delta_c$ at $t(z_c)=2t(z_{ta})$ can also
be calculated by evolving Eqs.~(\ref{x}) and (\ref{y}).  At early
time, the perturbation is linear
\begin{equation}
\frac{\rho_{cluster}}{\rho_b} = \left(\frac{x}{y}\right)^3\zeta 
\stackrel{x\rightarrow 0}{\longrightarrow} \delta_c(x) 
\end{equation}
Once $\delta_c$ is known at some $x_0$, then it is easily obtained
at an arbitrary time
\begin{equation} 
\delta_c(x)=\frac{g(x)}{g(x_0)}\delta_c(x_0)
\end{equation}
where $g$ is the growth factor that can be calculated by using 
Eq.~(\ref{g}).

\section{Growth Index in a Quintessense Universe}
\label{AppB}
The Q-component does not participate directly
in cluster
formation, but it alters the background cosmic evolution.  
The linear perturbation equation can be written as:
\begin{equation}  \label{linear}
\ddot{\delta}+2\frac{\dot{a}}{a}\dot{\delta}=4\pi G\rho \delta
\end{equation}
where $a$ is the scale factor of the Universe, dot means derivative
with respect to physical time $t$, $\delta=\delta \rho_m/\rho_m$, 
$\rho_m$ and $\delta \rho_m$ are the density and overdensity 
of the matter respectively.
The background evolution equations in a flat universe are:
\begin{eqnarray}
\label{friedman1}
\left(\frac{\dot{a}}{a}\right)^2=\frac{8\pi G}{3}
(\rho_m + \rho_Q)  \\
\label{friedman2}
2\frac{\ddot{a}}{a}+\left(\frac{\dot{a}}{a}\right)^2=
-8\pi G w \rho_Q 
\end{eqnarray}
where $\rho_Q$ is the energy density of the 
Q-component and $w\equiv p_Q/\rho_Q$ 
is the equation-of-state of the Q-component.
Now, we can define a matter energy density parameter 
$\Omega \equiv \Omega (a)$ so that:
\begin{equation}  \label{Omega}
H^2=\left(\frac{\dot{a}}{a}\right)^2=\frac{8\pi G}{3}\frac{\rho_m}{\Omega}
\end{equation}
 From Eqs.~(\ref{friedman1}), (\ref{friedman2}) and (\ref{Omega}),
we can get:
\begin{equation}
\label{dotaOmega}
\frac{\ddot{a}a}{\dot{a}^2}=-\frac 1 2 -\frac 3 2 w(1-\Omega)
\end{equation}
By using Eqs.~(\ref{friedman1}), (\ref{Omega}) and conservation of
stress energy $d(\rho a^3)=-pd(a^3)$:
\begin{equation}
\label{aOmega}
\frac{d\, \Omega}{d\, {\rm ln}\, a}=3w(1-\Omega)\Omega
\end{equation}
By using Eqs.~(\ref{linear}), (\ref{Omega}) and (\ref{dotaOmega})
we get:
\begin{equation}
\label{lineara}
\frac{d^2\, {\rm ln}\, \delta}{d\, {\rm ln}\, a^2}
+\left(\frac{d\, {\rm ln}\, \delta}{d\, {\rm ln}\, a}\right)^2
+\frac{d\, {\rm ln}\, \delta}{d\, {\rm ln}\, a}
\left[\frac 1 2 - \frac 3 2 w(1-\Omega)\right]=\frac 3 2 \Omega
\end{equation}
The growth index $f$ is defined as:
\begin{equation}
\label{fa}
f\equiv \frac{d\, {\rm ln}\, \delta}{d\, {\rm ln}\, a}
\end{equation}
By using Eqs.~(\ref{aOmega}) and (\ref{lineara}), we are able to
get the equation for $f$ in terms of $\Omega$:
\begin{equation}
\label{main}
3w\Omega(1-\Omega)\frac{d\, f}{d\, \Omega}+f\left[\frac 1 2 -
\frac 3 2 w(1-\Omega)\right] + f^2=\frac 3 2 \Omega
\end{equation}
Now, we introduce variable $\alpha$, so that $f \equiv 
\Omega ^{\alpha}$, and Eq.~(\ref{main}) becomes:
\begin{equation}
3w(1-\Omega)\Omega\ln{\Omega}\frac{d\, \alpha}{d\, \Omega}
-3w(\alpha -\frac 12)\Omega+\Omega^{\alpha}
-\frac 32\Omega^{1-\alpha} +3w\alpha -\frac 32 w +\frac 12 =0
\end{equation}
For slowly varying equation-of-state 
($ |d\, w/d\, \Omega| \ll 1/(1-\Omega ) $),  we shall get:
\begin{equation} \label{alpha_f}
\alpha=\frac{3}{5-\frac{w}{1-w}}
+\frac{3}{125}\frac{(1-w)(1-3w/2)}{(1-6w/5)^3}(1-\Omega)
+{\cal O}((1-\Omega)^2)
\end{equation}
By following a similar derivation, we found that 
Eq.~(\ref{alpha_f}) is valid for an open universe if
we set $w=-1/3$.   Hence,
$\alpha$ is weakly dependent on $\Omega_m$. The result is
    $\alpha \approx 6/11$
for $\Lambda$CDM (w=-1) and $\alpha \approx 4/7$ for OCDM.
\newpage
\begin{figure}
\epsfxsize=6 in \epsfbox{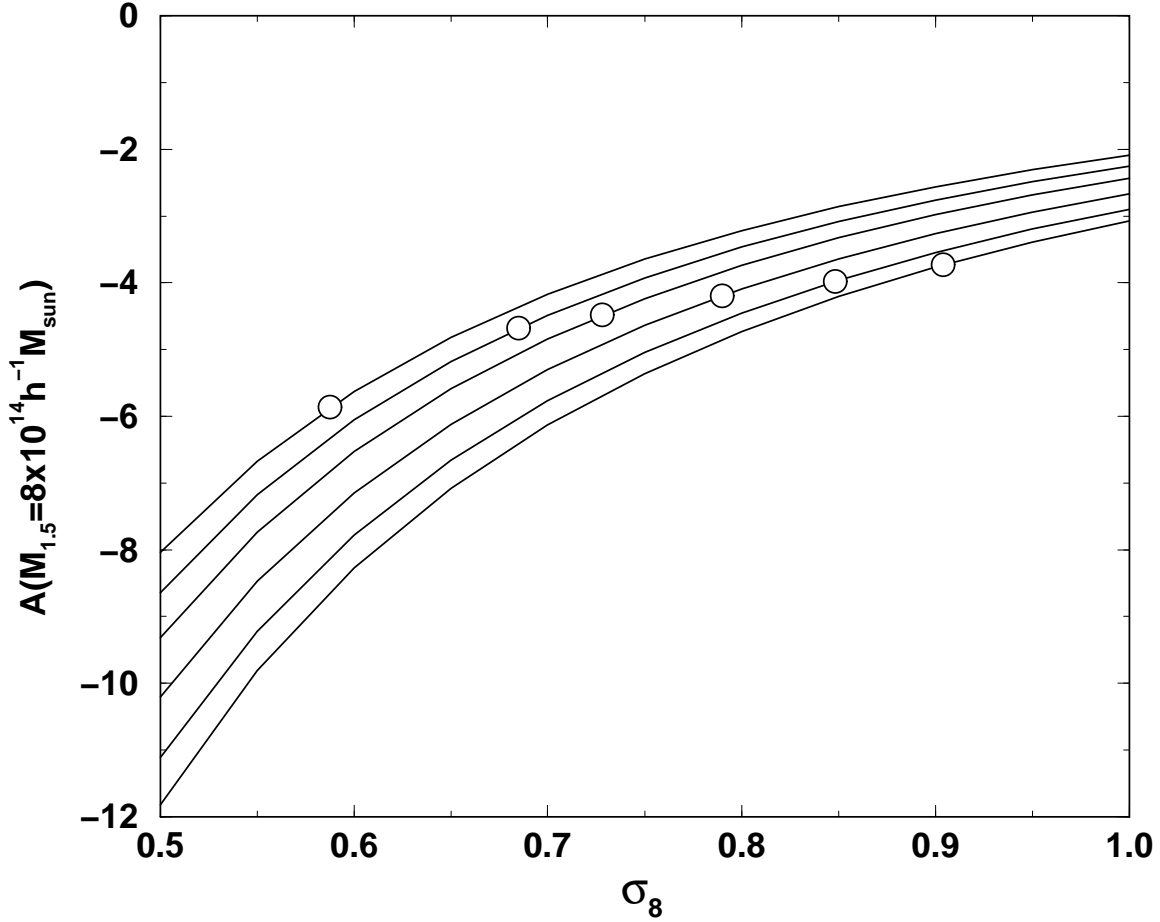}
\caption{The evolution index of cluster abundance 
for six models selected because they are the best-fit to 
the combination of COBE and cluster abundance (at $z=0$) constraints.
The model parameters  are
(from bottom to top):  
(1) $w=-1$, $\Omega_m=0.35$, $n_s=1$, $h=0.65$, $\Omega_b=0.047$; 
(2) $w=-5/6$, $\Omega_m=0.34$, $n_s=1$, $h=0.66$, $\Omega_b=0.046$; 
(3) $w=-2/3$, $\Omega_m=0.35$, $n_s=1$, $h=0.66$, $\Omega_b=0.046$; 
(4) $w=-1/2$, $\Omega_m=0.36$, $n_s=1$, $h=0.68$, $\Omega_b=0.043$; 
(5) $w=-1/3$, $\Omega_m=0.44$, $n_s=1$, $h=0.67$, $\Omega_b=0.045$; 
(6) $w=-1/6$, $\Omega_m=0.49$, $n_s=1.1$, $h=0.70$, $\Omega_b=0.042$.}
\end{figure}
\begin{figure}
\epsfxsize=6 in \epsfbox{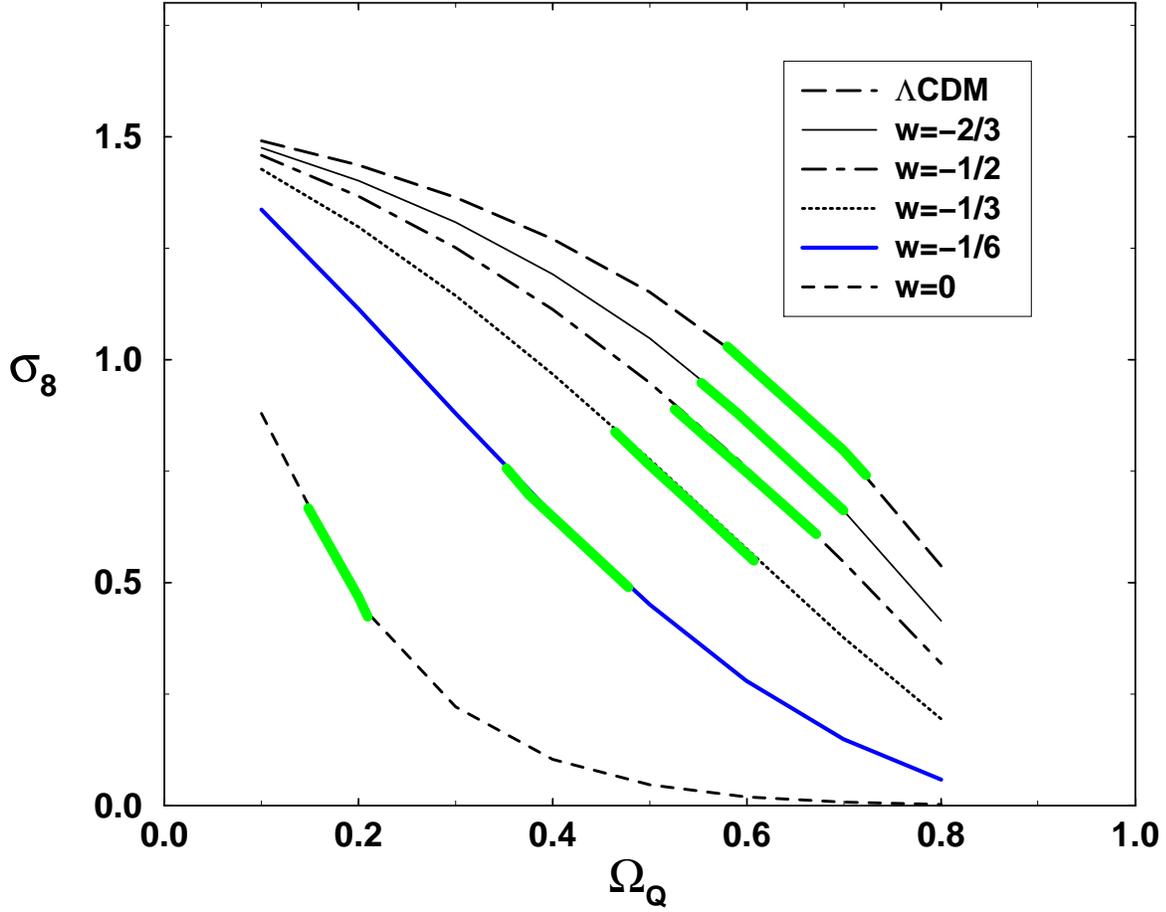}
\caption{COBE-normalized $\sigma_8$ as a function of $\Omega_Q$
for  six  constant $w$ models with $n_s=1$ and $h=0.65$.  They are
(from top to bottom):  
(1) $w=-1$;
(2) $w=-2/3$;
(3) $w=-1/2$;
(4) $w=-1/3$;
(5) $w=-1/6$;
(6) $w=0$.  
The highlighted regions indicate where the x-ray cluster
abundance constraints imposed by Eq.~(\ref{sigma8}) overlap the COBE
constraint. Best-fit models correspond to the overlap region.}
\end{figure}
\begin{figure}
\epsfxsize=6 in \epsfbox{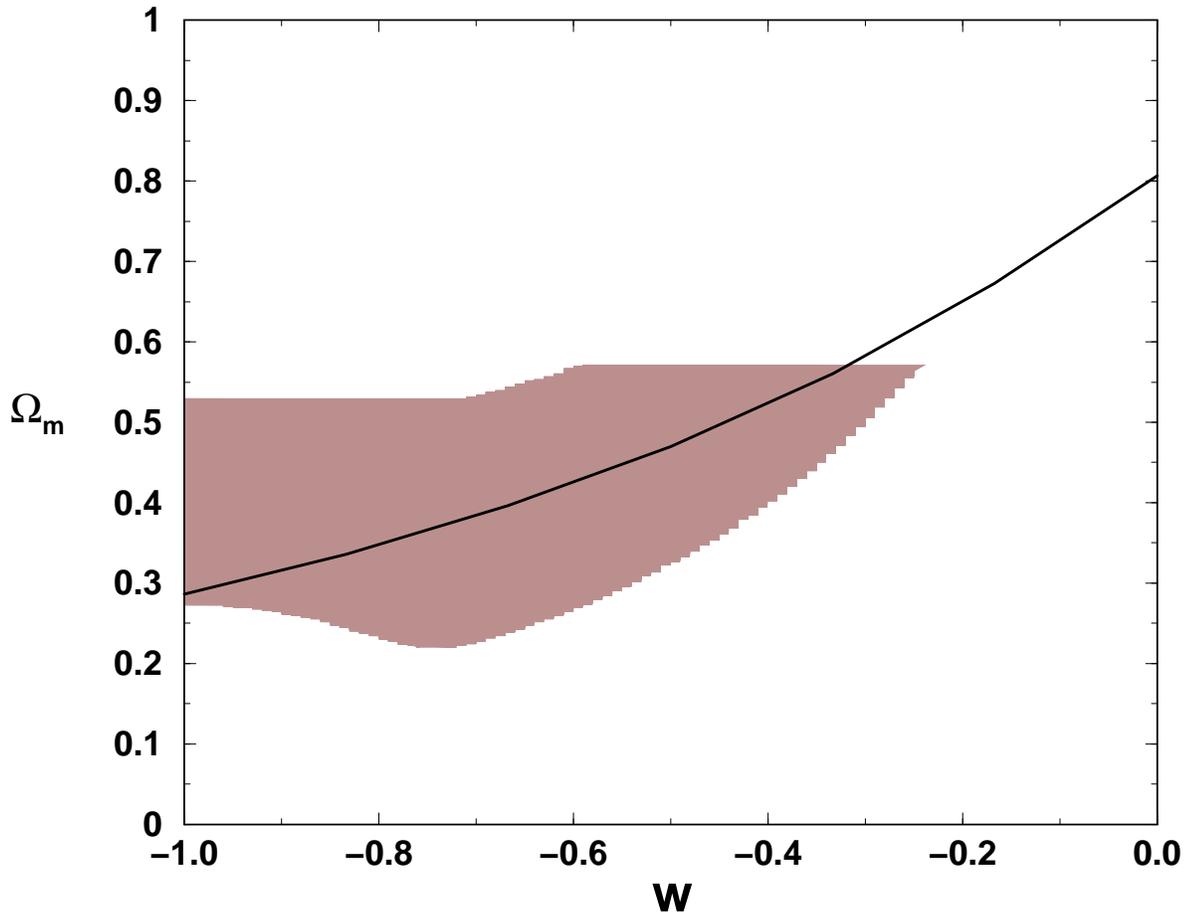}
\caption{Cosmic microwave background (CMB) anisotropy 
measurements, combined with other constraints on cosmological 
parameters, may be unable to distinguish among a family of $\Lambda$ and
quintessence models, as illustrated here.  
All models along the degeneracy curve shown in the figure produce
a temperature anisotropy power spectrum that is indistinguishable, given
cosmic variance uncertainty. In addition, the CMB can determine
 other parameters: for this illustration, we have assumed 
$\Omega_m h^2=1.5$, $\Omega_b h^2=0.02$ and $n_s=1$ (reasonable values).
Even if  these parameters are determined precisely and combined with other
observational constraints, there remains substantial
uncertainty (shaded region) that may not do much to discriminate among
the degenerate models, as illustrated here.
}
\end{figure}
\begin{figure}
\epsfxsize=6 in \epsfbox{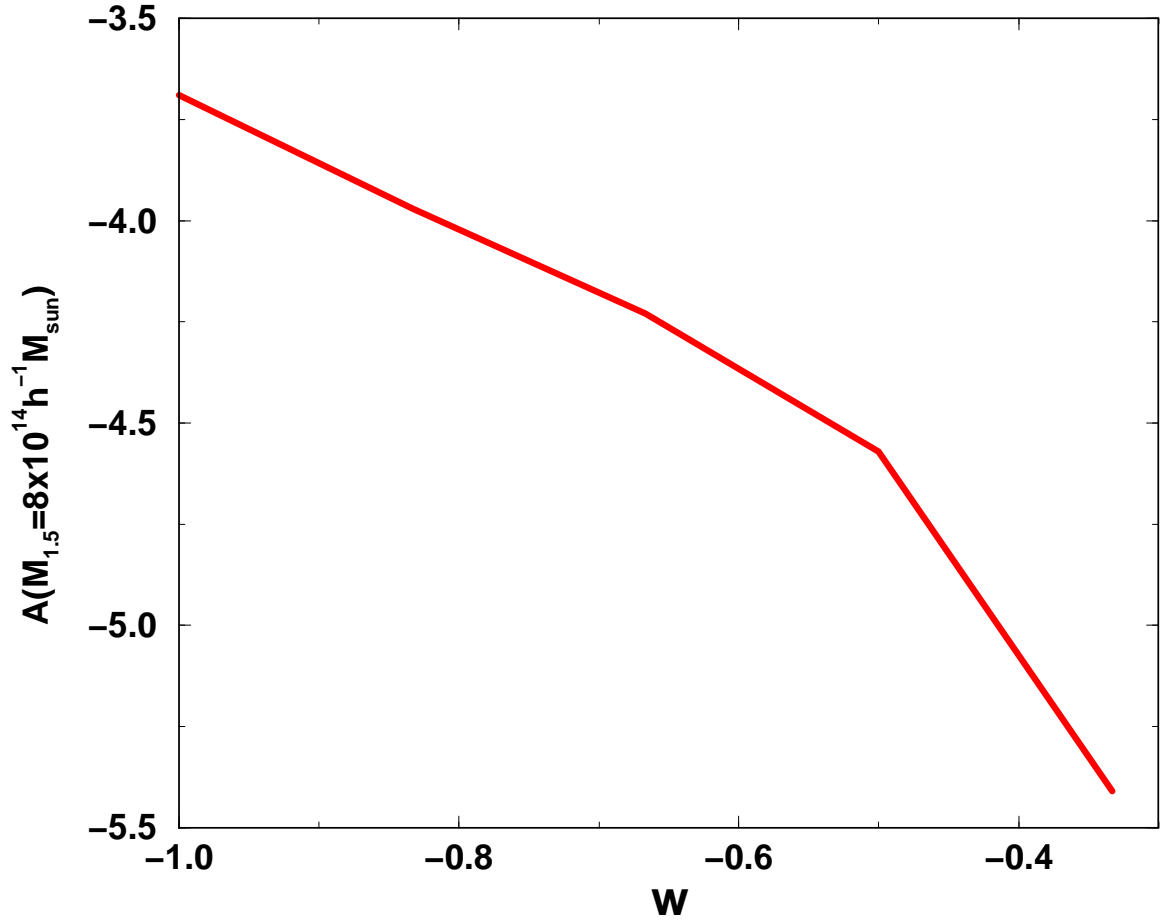}
\caption{The evolution of cluster abundance may 
 break down the degeneracy between $\Lambda$ and quintessence models 
illustrated in Fig.~3.
 $A(M_{1.5})$, the slope of log-abundance {\it vs.} redshift $z$,
 for the models along the degeneracy curve shown
in Figure~3. }
\end{figure}
\end{document}